\documentclass[twocolumn,showpacs,preprintnumbers,amsmath,amssymb]{revtex4}
\usepackage{graphicx}
\usepackage{bm}

\def\col#1#2{\left(\begin{array}{c} #1 \\[0mm] #2\end{array}\right)}

\newcommand{\ie}{{\it i.e.}}

\newcommand{\bfba}{\mathbb{A}\!\!\!\!\mathbb{A}}

\def\bra#1{ \langle #1\!\mid }
\def\ket#1{\mid\!#1\rangle}

\def\comm#1#2{\left[ #1,\,#2 \right]}

\begin{document}

\title{The effects of optically induced non-Abelian gauge field in cold atoms}

\author{Li-Hua Lu and You-Quan Li}
\affiliation{Zhejiang Institute of Modern Physics and Department of Physics,\\
Zhejiang University, Hangzhou 310027, P. R. China}

\begin{abstract}
 We show that $N-1$ degenerate dark states can be generated by
coupling $N$-fold degenerate ground states and a common excited
state with $N$ laser fields. Interferences between light waves
with different frequencies can produce laser fields with
time-dependent amplitudes, which can induce not only $U(N)$
non-Abelian vector fields but also the scalar ones for the
adiabatic motion of atoms in such laser fields. As an example, a
time-periodic gauge potential is produced by applying  specific
laser fields to a tripod system. Some features of the Landau
levels and the ground-state phase diagram of a rotating
Bose-Einstein condensate for a concrete gauge field are also
discussed.
\end{abstract}
\received{\today} 

\maketitle

\section{introduction}

In recent years, cooling atoms in laboratory is opening up a new
playground gathering various ingredients and distinct features
used to occur in condensed matter physics. Most of the prepared
systems are described by Hamiltonians formally identical to those
for electrons, which allows one to study the equivalent situations
under well defined controllable conditions. The electromagnetic
field, as a kind of gauge field, is known to be playing a
versatile role there. Like the emergence of the Coriolis force in
a rotating frame, rotating cold atoms trapped in a potential is a
useful way to provide an ``artificial" magnetic
field~\cite{RotatingRefs}. Although it had been noticed two
decades ago that the Berry phase~\cite{Berry,Wilczek} arising in
the adiabatic dynamics of quantum mechanical systems can be
regarded as a gauge field, the propagating of slow light in a
degenerate atomic gas was recently proposed to induce an effective
magnetic field for an electrically neutral
system~\cite{Juzeliunas04}. Laser beams with orbital angular
momenta have been suggested to induce either uniform or radially
dependent fields~\cite{Juzeliunas05}, which makes it possible to
discuss the quantum Hall effect and its incompressible nature in a
degenerate gas of atomic fermions trapped in a two dimensional
(2D) confinement~\cite{Ohberg05}. Alternatively, a method to
realize an artificial magnetic field for neutral atoms in a 2D
optical lattice was proposed~\cite{Jaksch}.

The realization of non-Abelian gauge fields in cold atomic systems
was theoretically suggested  very
recently~\cite{Sun,Dum,Osterloh,Ruseckas}. One
scheme~\cite{Osterloh} employs the laser-assisted-state-sensitive
tunnelling for atoms with multi-internal states, while another
one~\cite{Ruseckas} is based on doubly degenerate dark states
formed via a tripod-coupling system which is a developed scheme of
the dark state polariton ~\cite{Lukin}.
As we known, the spatial
components of gauge potentials were obtained directly whereas the
time component was introduced through a projection
procedure~\cite{Ruseckas} which is normally of higher order.

The possible emergence of non-Abelian gauge fields in various
branches of condensed matter physics has absorbed much attention
recently. For example, Rashba and Dresselhaus spin-orbit couplings
in certain semiconductors were perceived to be  non-Abelian gauge
fields ~\cite{Li}. It is therefore worthwhile to investigate the
emergence of non-Abelian gauge structures in neutral atoms
systematically.

In this paper, We show that both the non-Abelian vector potentials
and scalar ones can be induced in neutral atomic systems as long
as the amplitudes of the applied laser fields  are time-dependent.
Taking a tripod system as an example, we give time-periodic vector
and scalar gauge potentials using laser beams with particular
time-dependent amplitudes. We also discuss some concrete examples,
such as the features of the Landau levels  and the ground-state
phase diagram of a rotating Bose-Einstein condensate (BEC) system
in a definite gauge  potential. We find the  gauge potential can make
the phase diagram of a rotating BEC much richer even if the
expression of the gauge potential is  simple. In next section,
we consider a tripod system in the laser fields with
time-dependent amplitudes.
In section III, we consider Landau
levels and the ground-state phase diagram of a rotating system
in a definite gauge
potential. The main conclusions are summarized in  the last section.

\section{Model and general formulation}

We consider cold atomic ensembles moving in a radiation field which
couples resonantly the  atomic ground states of $N$-fold degeneracy
$\ket{\alpha}$ ($\alpha=1, 2,\cdots,N$) and a specific excited state
$\ket{e}$. In terms of the Rabi frequencies, we have the following
Hamiltonian in the representation of Hilbert subspace of internal
atomic levels
\begin{equation}\label{eq:inRadField}
\hat H'=-\hbar\sum_{\alpha=1}^N w^*_{\alpha}\ket{\alpha}\bra{e}
+\, \mathrm{ H.c.}.
\end{equation}
The Rabi frequency depends on the amplitude of the laser field
where the atom locates. The dressed state reads formally
$\displaystyle \ket{\psi}=(\mathcal{E}-\hat H
_0)^{-1}\hat{H}'\ket{\psi}$ with $\hat H_0$ the bare atomic
Hamiltonian, which gives rise to $\ket{\psi}=\sum_\alpha a_\alpha
\ket{\varphi^{}_\alpha}$  and
$\ket{\varphi^{}_\alpha}=\ket{\alpha}+(\mathcal{E} -\hat
H_0)^{-1}\hat P\hat{H}'
  \ket{\varphi^{}_\alpha}$
if the operator $\hat P$ projecting states outside of the subspace
for the degenerate ground states $\{\ket{\alpha}\}$ is introduced.
By iteration procedure one  reaches a formal expression up to any
expected order:
\begin{eqnarray*}\label{eq:perturbation}
\sum_{\alpha'}\bigl[
   \bra{\alpha}\!\hat H'\!\ket{\alpha'}
   +\bra{\alpha}\hat H'(\mathcal{E}-\hat H _0)^{-1} \hat P \hat H'\ket{\alpha'}
    + \cdots \bigr]a_{\alpha'}
    \\
=(\mathcal{E}-\mathcal{E}_0) a_\alpha, \hspace{38mm}
\end{eqnarray*}
where $\mathcal{E}_0$ is the unperturbated ground-state energy,
\ie, $\hat{H}_0\ket{\alpha}=\mathcal{E}_0\ket{\alpha}$. Since the
lowest nonvanishing perturbation for the Hamiltonian
(\ref{eq:inRadField}) is of the second order, we have
\begin{equation}\label{eq:secular}
\frac{1}{ \Delta\mathcal{ E}}\sum_{\alpha'}w^*_\alpha w_{\alpha'}
a_{\alpha'}=\mathcal{E}^{(1)} a_{\alpha},
\end{equation}
where  $\Delta\mathcal{E}$ is the energy difference between the
excited state $\ket{e}$ and the degenerate ground states
\{$\ket{\alpha}\}$, and
$\mathcal{E}=\mathcal{E}_0+\mathcal{E}^{(1)}$. By solving this
secular equation, we get $N-1$ degenerate eigenstates $\ket{d_i}$
with $\mathcal{E}^{(1)}=0$ and a single non-degenerate eigenstate
$\ket{s}$ with $\mathcal{E}^{(1)}=\sum_\alpha
\hbar^2|w_\alpha|^2/\Delta\mathcal{E}$, namely,
\begin{eqnarray}\label{eq:darkstates}
\ket{s} &=& \sum_{\alpha=1}^N \!\frac{w^*_\alpha}{\sqrt{|w_1|^2+|w_2|^2+\cdots+|w^{}_N|^2}}
   \ket{\alpha},
 \nonumber\\
\ket{d_i} &=& \sum_{\alpha=1}^{N}a^i_\alpha\ket{\alpha},
  \quad\quad (i=1,\, 2,\cdots, N\!-\!1).
\end{eqnarray}
The expanding coefficients in the state $\ket{d_i}$ yield
\begin{equation}\label{eq:coefficients}
\sum_\alpha w_\alpha a^i_\alpha=0,
\end{equation}
which implies that its solutions refer to $N-1$ orthogonal vectors
$\vec{a}=(a_1, a_2,\cdots,a^{}_N)$ perpendicular to a common vector
$\vec{w}=(w_1, w_2,\cdots, w^{}_N)$. Since $\bra{e}\hat H'
\ket{d_i}=0$, as a consequence of (\ref{eq:coefficients}), those
degenerate states $\ket{d_i}$ are actually dark states, which means
that the optically induced absorbtion  and emission occur
simultaneously.

The translational motion of the atom is described by the
time-dependent Schr\"{o}dinger equation $ i\hbar \partial_t
\ket{\Psi(\mathbf{r},t)} =\bigl[\mathbf{p}^2/2M + V(\mathbf{r})
\bigr]\ket{\Psi(\mathbf{r},t)}. $ Here $M$ denotes the mass of the
atoms and $V(\mathbf{r})$ an external  potential. The electric
dipole interaction $-\mathbf{D}\cdot\mathbf{E}(\mathbf{r},t)$
between atoms and the laser will not be directly  involved in the
Schr\"{o}dinger equation although the frequency $\omega$ of the
laser matches the energy difference between the aforementioned
ground and exited states, \ie, $\hbar\omega=\Delta\mathcal{E}$.
Whereas, we will see that radiation fields with  time-dependent
amplitudes will induce non-Abelian gauge fields in the
Schr\"{o}dinger equation for translational motions.

We apply a laser field with bi-frequency $\omega_+$ and $\omega_-$,
in stead of  a single frequency. Here $\omega=(\omega_+ +
\omega_-)/2$ matches for the occurrence of transition between the
atomic ground states and the excited state. The coherent
superposition of this two waves gives rise to a radiation field with
a time-dependent amplitude. One can find  that the Rabi frequencies
$w_\alpha$ are  time-dependent since they depend on the amplitude of
the radiation field. This brings about $(N-1)$-fold degenerate dark
states which depend on both time and spatial coordinates.

We apply such laser fields that the Born-Oppenheimer expansion is
applicable.
Substituting the state of the atom expanded in terms
of the dark states (as adiabatic basis)
\begin{equation*}\label{eq:psi}
\ket{\Psi(\mathbf{r},t)}
  =\sum_{i=1}^{N-1} \psi_i (\mathbf{r},t)\ket{d_i (\mathbf{r},t)}
\end{equation*}
into the Schr\"odinger equation for its translational motion,
we obtain that
\begin{eqnarray}\label{eq:schrodinger}
i\hbar\partial_t\psi = \Bigl[
   \frac{(\mathbf{p} - \bfba)^2}{2M}
     -\mathbb{A}_0 + V\Bigr]\psi,
\end{eqnarray}
where $\psi=(\psi_1,\,\psi_2,\cdots, \psi^{}_{N-1})^T$ and the
gauge potential $({\mathbb A}_0, \bfba)$ and $V$ are
matrix-valued. Their  matrix elements are given by
\begin{eqnarray}\label{eq:potentials}
\bfba_{ij}&=&i\hbar\bra{d_i(\mathbf{r},t)}
\nabla\ket{d_j(\mathbf{r},t)},
   \nonumber\\
(\mathbb{A}_0)_{ij}&=&i\hbar\bra{d_i(\mathbf{r},t)}\partial_t\ket{d_j(\mathbf{r},t)},\nonumber\\
V_{ij}&=&\bra{d_i(\mathbf{r},t)}V(\mathbf{r})\ket{d_j(\mathbf{r},t)},
\end{eqnarray}
clearly, $\mathbb{A}^\dagger_\mu=\mathbb{A}_\mu$, ($\mu=0, 1,
2,3$). Note that the scalar potential in Eq.~(\ref{eq:potentials})
differs from the one introduced in Ref.~\cite{Ruseckas} where the
scalar potential was obtained by means of a projection approach.
The Hamiltonian (\ref{eq:schrodinger}) is invariant under a local
$U(N-1)$ gauge transformation. The simplest non-Abelian case is a
tripod system which provides a U(2) gauge potential.

\section{Example of a tripod system}

As an example, we consider a tripod system which provides two
degenerate dark states. For convenience, we parameterize  the Rabi
frequencies $w_\alpha$ with angle and phase variables according to
$w_1=w\sin\theta\cos\phi e^{iS_1}$, $w_2=w\sin\theta\sin\phi
e^{iS_2}$, $w_3=w\cos\theta e^{iS_3}$, where
$w=\sqrt{|w_1|^2+|w_2|^2+|w_3|^2}$, $\tan\phi={|w_2|}/{|w_1|}$, and
$\tan\theta=\sqrt{({|w_1|^2+|w_2|^2})/{|w_3|^2}}$.  According to
Eq.(\ref{eq:coefficients}), the two dark states are given by
\begin{eqnarray}
\ket{d_1} &=& e^{-i\chi_1} \bigl(\sin\phi
e^{-iS_1}\ket{1}-\cos\phi e^{-iS_2}\ket{2}\bigr),\nonumber\\
\ket{d_2} &=& e^{-i\chi_2}\bigl(\cos\theta\cos\phi
e^{-iS_1}\ket{1}+\cos\theta\sin\phi e^{-iS_2}\ket{2}\nonumber\\
&&-\sin\theta e^{-iS_3}\ket{3}\bigr),
\end{eqnarray}
where $\chi_1$ and $\chi_2$ are arbitrary phase variables. The
matrix elements of the gauge potential $(\mathbb{A}_0, \bfba )$
are calculated easily, which yield
\begin{eqnarray}\label{eq:vector A}
\bfba_{11} &=& \hbar\bigl(\nabla (S_1+\chi_1) +\cos^2\phi\nabla
S_{21}\bigr),\nonumber\\
\bfba_{12} &=& \hbar\cos\theta
\bigl(\frac{1}{2}\sin(2\phi)\nabla S_{12}-i\nabla\phi\bigr)e^{i(\chi_1-\chi_2)},
  \nonumber\\
\bfba_{22} &=& \hbar\bigl(\cos^2\theta\cos^2\phi\nabla
S_{12}+\cos^2\theta\nabla S_{23}\nonumber\\
&&+\nabla (S_3+\chi_2)\bigr) ,
\end{eqnarray}
and
\begin{eqnarray}\label{eq:scalar A}
(\mathbb{A}_0)_{11} &=& \hbar\bigl(\partial_t (S_1+\chi_1)
+\cos^2\phi\partial_tS_{21}\bigr),\nonumber\\
(\mathbb{A}_0)_{12} &=& \hbar\cos\theta
\bigl(\frac{1}{2}\sin(2\phi)\partial_t S_{12}-i\partial_t\phi\bigr)
    e^{i(\chi_1-\chi_2)},\nonumber\\
(\mathbb{A}_0)_{22} &=&
\hbar\bigl(\cos^2\theta\cos^2\phi\partial_t
S_{12}+\cos^2\theta\partial_tS_{23}\nonumber\\
&&+\partial_t (S_3+\chi_2)\bigr),
\end{eqnarray}
where $S_{ij}=S_i-S_j$. The gauge potentials $(\mathbb{A}_0, \bfba
)$ corresponding to  different $\chi_1$ and $\chi_2$ are related by
a local $U(2)$ gauge transformation.

We now construct specific radiation fields which lead to a
time-periodic  gauge field. For this purpose, we consider  two
copropagating laser beams in the $z$-direction with slightly
different frequencies, whose electric fields read
\begin{eqnarray}
\displaystyle \mathbf{E}_+&=&\frac{1}{2}E_0
e^{i(k_+z-\omega^{}_{+}t+\vartheta_+)}\hat{e}_z,
  \nonumber\\
\displaystyle\mathbf{E}_-&=&\frac{1}{2}E_0e^{i(k_-z-\omega^{}_-t+\vartheta_-)}\hat{e}_z,
   \nonumber
\end{eqnarray}
where $k_\pm$   are the wave vectors of the two fields,
$\omega^{}_\pm$ the frequencies and $\vartheta_\pm$ the initial
phases. The coherent superposition of the two fields gives rise to
\begin{equation}
\mathbf{E}_{super}=E_0\cos(\Delta kz-\Delta \omega t +
\Delta\vartheta) e^{i(k'z+\vartheta)}e^{-i\omega
t}\hat{e}_z,\nonumber
\end{equation}
where $k_+ -k_-=2\Delta k$, $k_+ +k_-=2k'$, $\omega^{}_+
-\omega^{}_-=2\Delta \omega$, $\omega^{}_+ +\omega^{}_-=2\omega$,
$\vartheta_+ -\vartheta_-=2\Delta\vartheta$ and $\vartheta_+
+\vartheta_-=2\vartheta$. Fixing  the initial phase-difference
$2\Delta\vartheta$ on zero, we can get a field
$\mathbf{E}_{super}$
  with the complex amplitude being
\begin{equation}\label{eq:complex}
U_1=E_0\cos(\Delta k z-\Delta\omega t)e^{i(k'z+\vartheta)}.
\end{equation}
The Rabi frequencies corresponding to the above field is
\begin{equation}\label{eq:omega1}
w_1=w_0\cos(\Delta k z-\Delta\omega t)e^{i(k'z+\vartheta)},
\end{equation}
where  $\hbar w_0=-E_0{(D_z)_{1e}}$ and $(D_z)_{1e}=\bra eD_z\ket
1$ with $D_z$ being the $z$-component of the electric-dipole
moment of the atom. Additionally, two copropagating and circularly
polarized fields with opposite orbital angular momenta are applied.
The Rabi frequencies of those fields are
\begin{equation}\label{eq:omega3}
w_{2,3}=w_0 \rho e^{i(kz\pm\varphi)},
\end{equation}
in cylindrical coordinates.

The gauge potential corresponding to the above fields
Eq.({\ref{eq:omega1}) and Eq.(\ref{eq:omega3}) can be calculated
from Eq.(\ref{eq:vector A}) and Eq.(\ref{eq:scalar A}),
 \begin{widetext}
 \begin{eqnarray}
\bfba &=& \hat{e}_\rho\hbar\cos\theta\left(%
\begin{array}{cc}
  0 & -i \\
  i & 0 \\
\end{array}%
\right)\frac{\cos(\Delta kz-\Delta\omega
t)}{\cos^2(\Delta kz-\Delta\omega t)+\rho^2}+\hat{e}_\phi\frac{\hbar}{\rho}\left(%
\begin{array}{cc}
  \cos^2\phi & \displaystyle-\frac{\cos\theta\sin(2\phi)}{2} \\
  \displaystyle-\frac{\cos\theta\sin(2\phi)}{2} & \quad  \cos^2\theta\sin^2\phi-\sin^2\theta \\
\end{array}%
\right)\nonumber\\[2mm]
&&+\hat{e}_z\hbar\left(%
\begin{array}{cc}
  k'\sin^2\phi+k\cos^2{\phi} &\displaystyle \cos\theta\Bigl[\frac{\sin(2\phi)(k'-k)}{2}-i\frac{\Delta k\rho  \sin(\Delta k z-\Delta\omega t)}{\cos^2(\Delta kz-\Delta\omega t)+\rho^2}\Bigr]
  \\
 \displaystyle\cos\theta\Bigl[\frac{\sin(2\phi)(k'-k)}{2}+i\frac{\Delta k\rho  \sin(\Delta k z-\Delta\omega t)}{\cos^2(\Delta kz-\Delta\omega t)+\rho^2}\Bigr]& k+\cos^2\theta\cos^2{\phi}(k'-k) \\
\end{array}%
\right),\nonumber\\[3mm]
\mathbb{A}_0&=&\hbar\cos\theta\left(%
\begin{array}{cc}
  0 & i \\
  -i & 0 \\
\end{array}%
\right)
\frac{\Delta\omega\rho\sin(\Delta kz-\Delta \omega t)}
    {\cos^2(\Delta kz-\Delta\omega t)+\rho^2},
\end{eqnarray}
\end{widetext}
as long as $\chi_1$ and $\chi_2$ are chosen as zero. Here
$\cos\phi=\cos(\Delta kz-\Delta\omega
t)\big/\sqrt{\rho^2+\cos^2(\Delta kz-\Delta\omega t)}$ and
$\cos\theta=\rho\big/\sqrt{2\rho^2+\cos^2(\Delta kz-\Delta\omega
t)}$. We can easily find that the gauge potential
($\mathbb{A}_0$,$\bfba $ ) is periodic in time with a periodicity
of $2\pi/\Delta \omega$.

\section{Example of a rotating system  }

We now take a two-dimensional rotating cold atomic system as an
example to investigate the effect of the non-Abelian gauge fields.
For the rotating cold atomic system, the Hamiltonian is written as
\begin{equation}\label{eq:hbe}
 H=\frac{\mathbf{p}^2}{2M}+\frac{1}{2}M\omega_\bot^2 x^2
 +\frac{1}{2}M\omega_\bot^2 y^2-\Omega \hat{L}(z)+H_{int},
 \end{equation}
in a rotating frame. Here $\omega_\bot$ denotes the frequency of
the trapping potential, $\Omega$ the rotating frequency of the
system,  $\hat{L}(z)=(\mathbf{r}\times \mathbf{p})_z$  the
conventional angular momentum operator and $H_{int}$ the
interaction between atoms. The application of laser fields
\begin{eqnarray}
w_1 &=& \sqrt{\frac{\hbar k-2\gamma x}{2\hbar k}}
e^{i(-ky+\delta_1)},\nonumber\\
 w_2 &=& \sqrt{\frac{\hbar k-2\gamma x}{2\hbar k}}
e^{i(-ky+\delta_2)},\nonumber\\
w_3 &=& \sqrt{ \frac{2\gamma x}{\hbar k}}
e^{i(-kz+\delta_3)},\nonumber
\end{eqnarray}
with $\delta_1$, $\delta_2$ and $\delta_3$ being the initial
phases  and $k$ the wave vector of the laser fields, introduces
the non-Abelian gauge potential $\bfba$ into the above Hamiltonian
for rotating Bose-Einstein condensates, accordingly
\begin{eqnarray}\label{eq:adh}
H&=&\frac{(\mathbf{p}-\bfba)^2}{2M}+\frac{1}{2}M\omega_\bot^2 x^2
 +\frac{1}{2}M\omega_\bot^2 y^2\nonumber\\
 &&-\Omega\cdot\bigl(\mathbf{r}\times(\mathbf{p}-\bfba)\bigr)_z+H_{int}.
\end{eqnarray}
Here $\mathbf{p}=(p_x, p_y)$, $\bfba=(\mathbb{A}_x,
\mathbb{A}_y)$, $\mathbb{A}_x = -\frac{1}{2}\gamma y
\hat{\tau}_z+\frac{1}{2}\gamma yI$ and
$\mathbb{A}_y=\frac{1}{2}\gamma x \hat{\tau}_z - \frac{1}{2}\gamma
xI$ with $\hat{\tau}_z$ being the $z$-component of Pauli matrix.
Collecting the harmonic trapping potential terms
into the quadratic term $(\mathbf{p}-\bfba)^2\big/(2M)$,
we can write Eq.~(\ref{eq:adh}) as
\begin{equation}\label{eq:Hh}
H = \frac{1}{2M}(\mathbf{p}-\mathbf{A}- \bfba)^2
   +(\omega^{}_\bot- \Omega) \mathbb{L}(z) + H_{int},
\end{equation}
where $\mathbf{A}=(A_x, A_y)$,
 $A_x=-M\omega^{}_\bot y$, $A_y=M\omega^{}_\bot x$ and
$\mathbb{L}(z)=\hat{L}(z)-(\mathbf{r}\times
\bfba)_z=\bigl(\mathbf{r}\times (\mathbf{p}-\bfba)\bigr)_z$.

In the limit of $\Omega\rightarrow\omega_\bot$,
the Hamiltonian is formally the same as that for a charged particle in the
background of conventional  and   $U(2)$ magnetic fields.
In this case, $(A_x,\,A_y)$ becomes $(-M\Omega y,\,M\Omega x)$,
and the corresponding magnetic field reads $B_z=2M\Omega$.
The $i$-component
of the dynamical momentum is given by
\begin{equation}
 \mathbb{P}_i = \frac{M}{i\hbar} \comm{r_i}{\hat{H}}
  =\hat{p}^{}_i-A_i-\mathbb{A}_i.
\end{equation}
The commutator between different components of the dynamical
momentum is shown to obey
\begin{equation}
\comm{\mathbb{P}_x}{\mathbb{P}_y}
=i\hbar(B_z+\mathbb{B}_z)=i\hbar\left(%
\begin{array}{cc}
  B_1 & 0 \\
  0 & B_2\\
\end{array}%
\right),
\end{equation}
where $\mathbb{B}_z= \gamma \hat{\tau}_z -\gamma I$ refers to
the $U(2)$ magnetic field which is defined by the corresponding
$U(2)$ gauge potential.
In general,
$\mathbb{B}_i=\frac{1}{2}\epsilon_{ijk}(\partial _j
\mathbb{A}_k-\partial _k \mathbb{A}_j
-\frac{i}{\hbar}[\mathbb{A}_j,\mathbb{A}_k])$. Clearly, different
components of the dynamical momentum are noncommutative as long as
either the U(2) magnetic field $\mathbb{B}_z$ or the conventional
magnetic field $B_z$ is not zero. Following the strategy of
algebraic approach to Landau levels, we introduce the
pseudomomentum whose $i$-component is
$\mathbb{K}_i=\hat{p}_i+A_i+\mathbb{A}_i$. The Hamiltonian (in the
absence of interactions between atoms) and the angular momentum
operator can be written as
\begin{eqnarray}
\hat{H} &=& \frac{1}{2M}(\mathbb{P}^2_x+\mathbb{P}^2_y),
  \nonumber\\
\hat{L}_z&=&\frac{1}{2}(B_z+\mathbb{B}_z)^{-1}(\mathbb{K}^2_x+\mathbb{K}^2_y-\mathbb{P}^2_x-\mathbb{P}^2_y),
\end{eqnarray}
where $(B_z+\mathbb{B}_z)(B_z+\mathbb{B}_z)^{-1}=I$.
We can thus define two sets of creation and annihilation operators
with the above dynamical momentum and the  pseudomomentum.
Each operator has two species
$$
\hat a =\left(%
\begin{array}{cc}
  \hat{a}_1 & 0 \\
 0 & \hat{a}_2\\
\end{array}%
\right)=(\mathbb{P}_x + i\mathbb{P}_y)\left(%
\begin{array}{cc}
  \displaystyle\frac{1}{\sqrt{2\hbar B_1}} & 0 \\
  0 & \displaystyle\frac{1}{\sqrt{2\hbar B_2}} \\
\end{array}%
\right),$$
 $$ \hat b =\left(%
\begin{array}{cc}
  \hat{b}_1 & 0\\
  0 & \hat{b}_2 \\
\end{array}%
\right)=(\mathbb{K}_x - i\mathbb{K}_y)\left(%
\begin{array}{cc}
  \displaystyle\frac{1}{\sqrt{2\hbar B_1}} & 0 \\
  0 & \displaystyle\frac{1}{\sqrt{2\hbar B_2}} \\
\end{array}%
\right).
$$
In terms of those creation and annihilation operators,
the Hamiltonian $\hat{H}=(\mathbb{P}_x^2+\mathbb{P}_y^2)/2M$ is expressed as
\begin{equation}
\hat{H}=\sum_{i=1,2}\hbar\omega_i \hat{a}^+_i\hat{a}_i,
\end{equation}
whose eigenfunctions and eigenvalues  are $(\ket{n_1, m_1},
\ket{n_2, m_2})^T$, and $E_{n_1
n_2}=\hbar\omega_1(2n_1+1)+\hbar\omega_2(2n_2+1)$,
 respectively, where
$\omega_1=\omega_\bot$  and $\omega_2=\omega_\bot-\gamma/M$. Here
the states $|n_1,m_1\rangle$ and $|n_2,m_2\rangle$ are Landau levels
with $n_1,m_1$ and $n_2,m_2$ being integers. The integers $n_1$ and
$n_2$ are the Landau level indices, and $m_1$ and $m_2$ label the
degenerate states within the same Landau level.

In  the following discussion, we assume that the system is in the
lowest Landau level (LLL), \ie,  $n_1,\, n_2 = 0$. The assumption
is valid in the limit of $\Omega\sim\omega_\bot$. The wave
function of the system is written as
\begin{eqnarray}\label{eq:wave}
\Psi&=& \col{\Psi_1}{\Psi_2},
   \nonumber\\
\Psi_{1} &=& \sum_{m_{1}}c_{m_{1}}r^{m_{1}}e_{}^{im_{1}\varphi }
\displaystyle
   e_{}^{-\frac{r^2}{2a_{1}^2 } },\nonumber\\
   \Psi_{2} &=& \sum_{m_{2}}c_{m_{2}}r^{m_{2}}e_{}^{im_{2}\varphi }
\displaystyle
   e_{}^{-\frac{r^2}{2a_{2}^2 } },
\end{eqnarray}
where $a_1=\sqrt{\hbar/(M\omega_1)}$, and
$a_2=\sqrt{\hbar/(M\omega_2)}$. The expectation  value of
$\mathbb{L}(z)$ for the wave function (\ref{eq:wave}) is given by
\begin{eqnarray}\label{eq:L}
\langle\mathbb{L}(z)\rangle&=&\hbar\int[(r/a_1)^2-1]|\Psi_1|^2
d{\bf r}\nonumber\\ &&+\hbar\int[(r/a_2)^2-1]|\Psi_2|^2 d{\bf r}\nonumber\\
  &&   + \gamma\int r^2|\Psi_2|^2 d{\bf r}.
\end{eqnarray}
Then the expectation  value of Eq.~(\ref{eq:Hh}) can be evaluated
\begin{eqnarray}\label{eq:gh}
\langle\hat{H}\rangle
&=&(\omega_\bot-\Omega)M\omega_\bot\bigl(\langle
r^2\rangle^{}_1+\langle
r^2\rangle^{}_2\bigr)+ 2\hbar\Omega+H_{int}, \nonumber\\
H_{int}&=&\int
d^2r\Bigl[\frac{1}{2}g_1|\Psi_1|^4+\frac{1}{2}g_2|\Psi_2|^4
 \nonumber\\
&& \hspace{6mm}   + g_{12}|\Psi_1|^2|\Psi_2|^2\Bigr],
\end{eqnarray}
where $\langle r^2\rangle^{}_1=\int r^2|\Psi_1|^2 d {\bf r}$ and
$\langle r^2\rangle^{}_2=\int r^2|\Psi_2|^2 d{\bf  r}$. The
strengths of the effective two-body interactions are
$g_1=4\pi\hbar^2 a_{s1}N\big/ZM$, $g_2=4\pi\hbar^2
a_{s2}N\big/ZM$, and $g_{12}=4\pi\hbar^2 a_{s12}N\big/ZM$. Here
the variation of the trapping potential along the $z$-axis is
neglected, and the density of atoms along $z$-axis is assumed to
be a constant $N/Z$.
\begin{figure}[tbph]
\includegraphics[width=60mm]{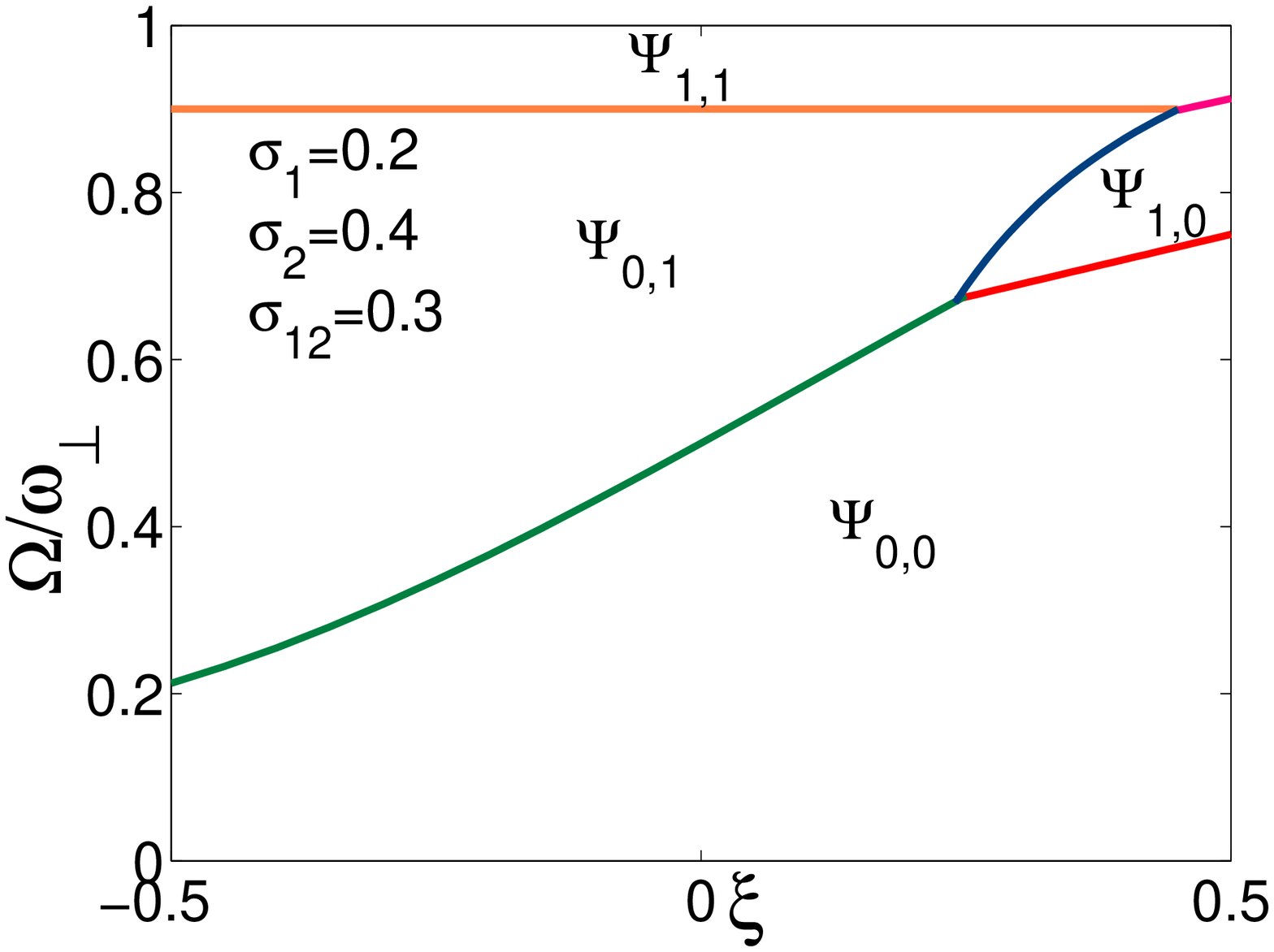}(a)
\includegraphics[width=60mm]{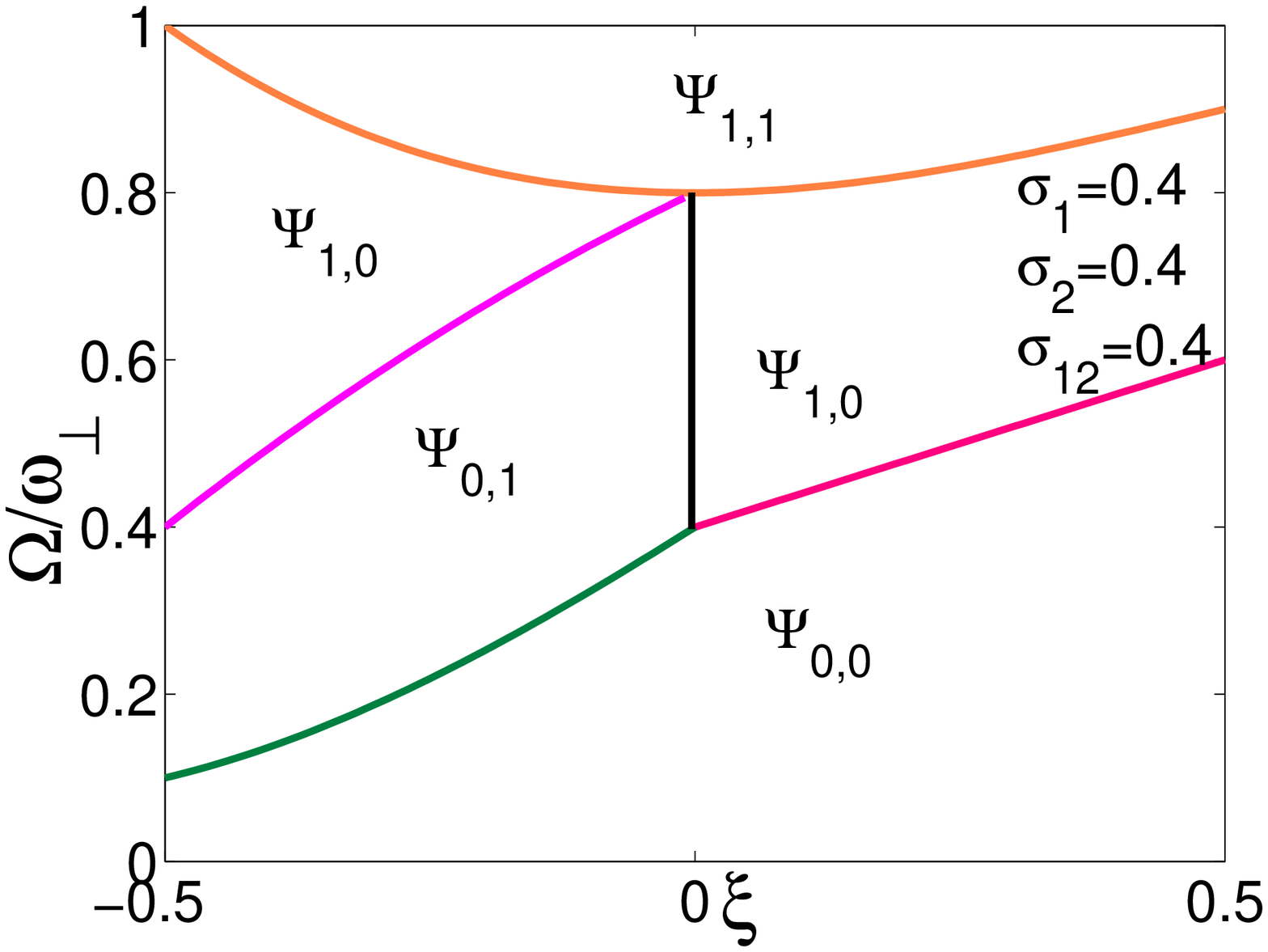}(b)
\caption{\label{fig:phim} (color online) Ground-state phase
diagrams with different choices of the parameters for a rotating
BEC in the presence of gauge potentials in the
$\xi$-$\Omega/\omega_\bot$ plane. The boundaries separating
different phases are plotted. $\sigma_i$, $\sigma_{12}$,
$\Omega/\omega_{\bot}$ and $\xi$ are all dimensionless
quantities.}
\end{figure}

 For sufficiently weak interactions,
 the energy of
atoms in the rotating frame is minimized when the wave functions
$\Psi_1$, $\Psi_2$ are composed of only one component,
respectively, \ie, $\Psi_{m_1,m_2}= $ $(r^{m_1} e^{im_1\varphi}
e^{-r^2/2a_1^2},r^{m_2}e^{im_2\varphi}e^{-r^2/2a_2^2})^T$~\cite{Jackson}.
In this case the energy of the system becomes
\begin{eqnarray}\label{eq:energy}
\mathcal{E}_{m_1,m_2}&=&
(1-\frac{\Omega}{\omega_\bot})\Bigl[m_1+1+\frac{m_2+1}{1-\xi}\Bigr]\nonumber\\
&&+\sigma_1\frac{(2m_1)!}{2^{2m_1}(m_1!)^2}
+\sigma_2(1-\xi)\frac{(2m_2)!}{2^{2m_2}(m_2!)^2}\nonumber\\
&&+\sigma_{12}\frac{(m_1+m_2)!(1-\xi)^{(m_2+1)}}{2^{m_1+m_2-1}m_1!m_2!},
\end{eqnarray}
in unit of $\hbar\omega_\bot$. Here $\xi\equiv\gamma\big/(
M\omega_\bot)$, $\sigma_i \equiv a_{si}N\big/Z$ and
$\sigma^{}_{12}\equiv a_{s12}N\big/Z$. For definite values of
$\Omega$ and $\xi$, the energy of the system depends on the
quantum numbers $m_1$ and $m_2$. The optimal values of $m_1$ and
$m_2$ corresponding to the ground state can be determined by
minimizing the energy Eq.~(\ref{eq:energy}).

From Eq.~(\ref{eq:energy}) we can see that $\mathcal{E}_{1,0}$ and
$\mathcal{E}_{0,1}$ become lower than $\mathcal{E}_{0,0}$ at
critical frequencies of rotation given by
\begin{eqnarray}
\frac{\Omega_{1,0}}{\omega_\bot} &=& 1-\frac{\sigma_1}{2}-(1-\xi)\sigma_{12},\nonumber\\
\frac{\Omega_{0,1}}{\omega_\bot}&=&1-\frac{(1-\xi)^2\sigma_2}{2}-\sigma_{12}(1+\xi)(1-\xi)^2,
\end{eqnarray}
respectively. The vortex begins to be created in one component of
the system when the frequency of the rotation reaches  the minimal
value of \{$\Omega_{1,0}$, $\Omega_{0,1}$\}. For some values of
$\xi$, vortices can be created in both components of the system
when $\Omega$ reaches some critical value $\Omega_{m_1,m_2}$.
Thus the state $\Psi_{m_1,m_2}$ with $m_1$ and $m_2$ being both
larger than zero becomes more favorable in energy for
$\Omega\geq\Omega_{m_1,m_2}$. This critical frequency
$\Omega_{m_1,m_2}$ can be calculated from the two conditions,
\begin{equation}
\mathcal{E}_{m_1,m_2}-\mathcal{E}_{m_1-1,m_2}=0,\quad
\mathcal{E}_{m_1,m_2}-\mathcal{E}_{m_1,m_2-1}=0\nonumber.
\end{equation}
Thus we can plot out the ground-state phase diagram of the
rotating BEC  in which the boundaries between different phases are
obtained by the relation
$\mathcal{E}_{m_1,m_2}-\mathcal{E}_{m'_1,m'_2}=0$.
Fig.~\ref{fig:phim} is plotted for two different choices of the
parameters where the $\sigma$'s are small in order to satisfy the
assumption of weak interactions. One can see that the shapes of
the phase diagram for the two parameter choices change distinctly.
This is due to the differences of the scattering lengths can break
the spatial symmetry of the ground state. From
Fig.~\ref{fig:phim}, we can find the angular momentum of the
system depends not only on the rotating frequency $\Omega$ but
also on the strength of the gauge potential for  given
interactions. Discontinuous transitions may be observed through
changing the value of $\xi$ even if  $\Omega$ is a constant. The
presence of gauge potentials makes the phase diagram of the system
much richer.

\begin{figure}[tbph]
\includegraphics[width=80mm]{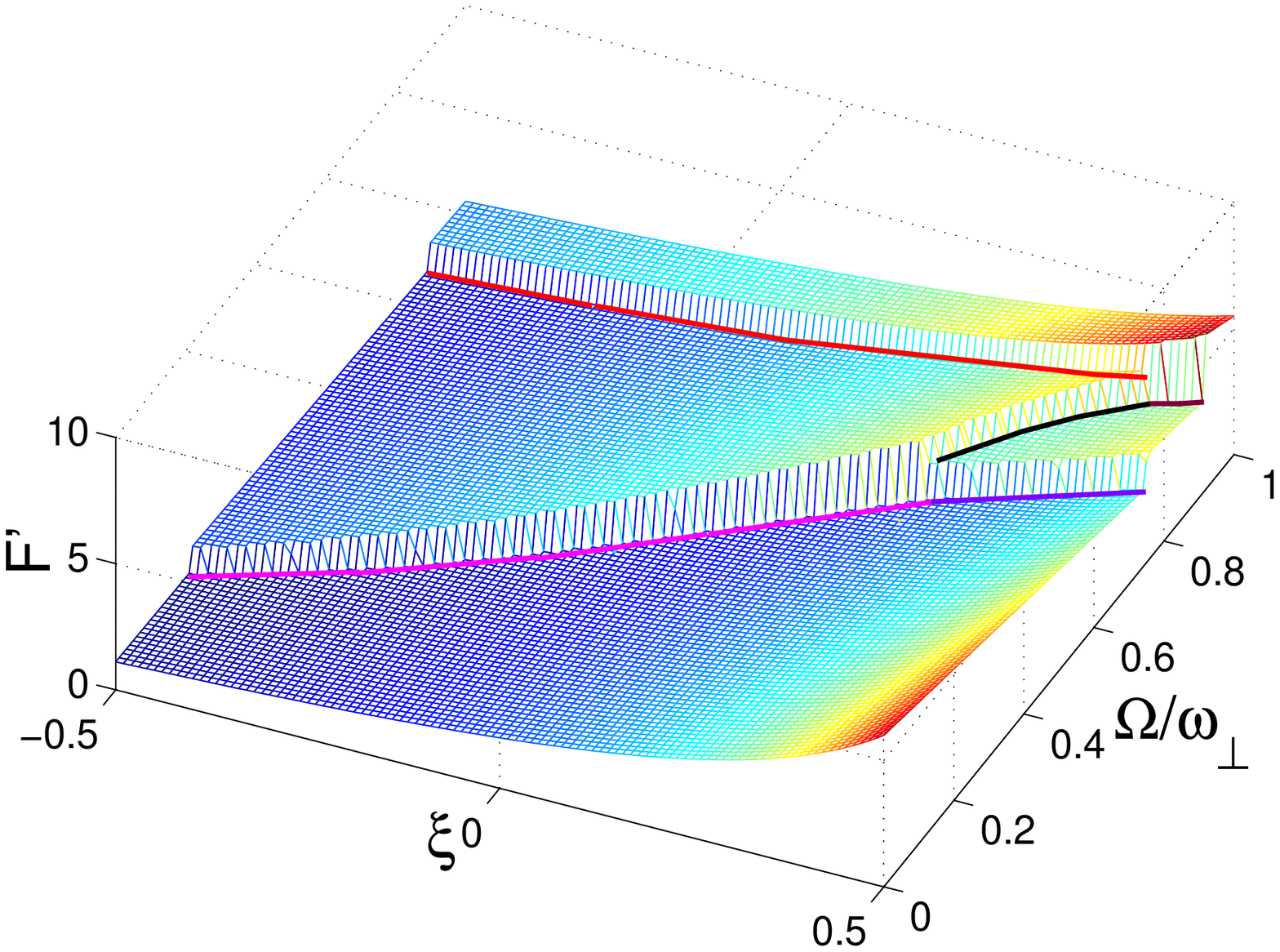}(a)
\includegraphics[width=80mm]{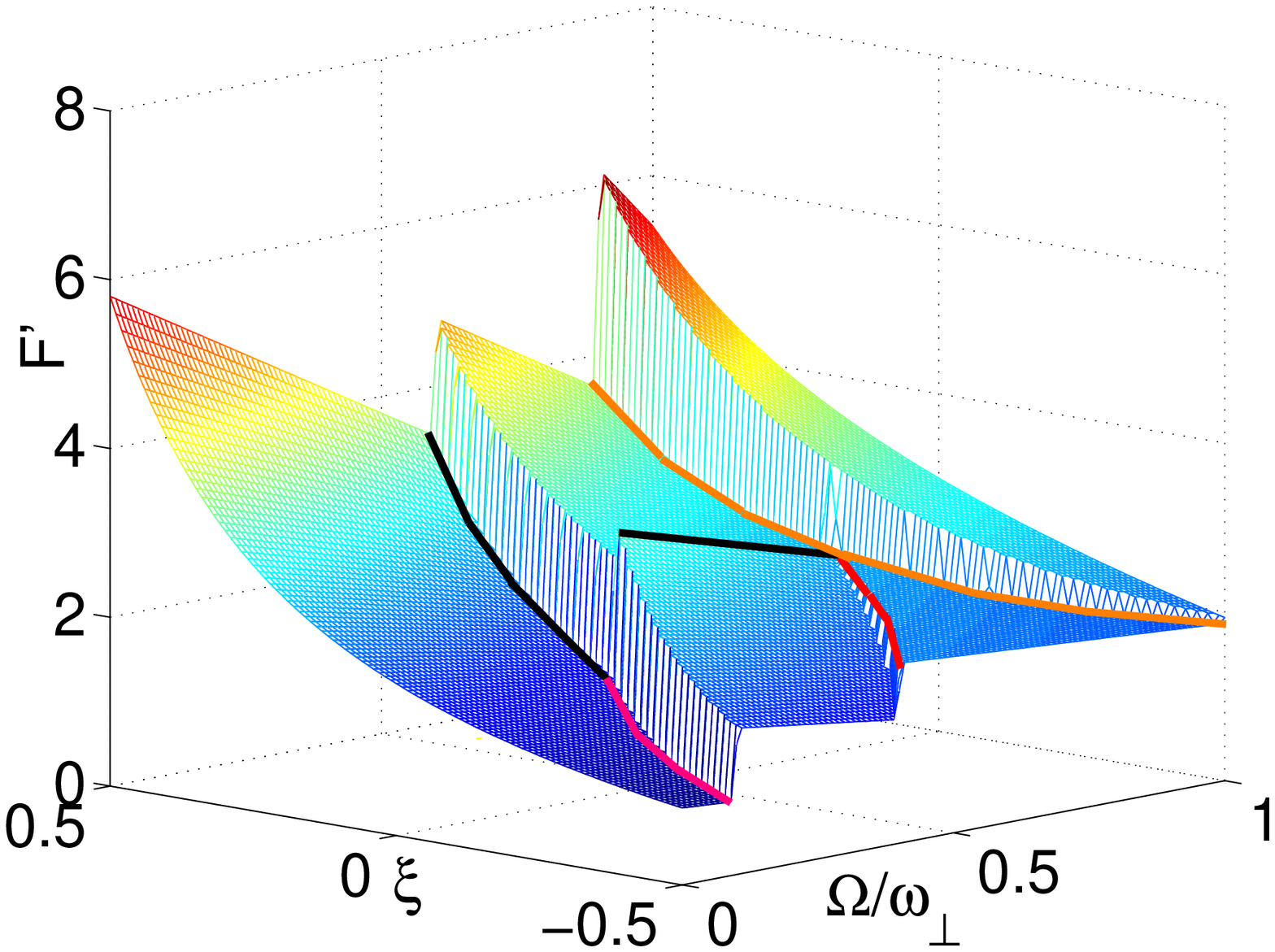}(b)
\caption{\label{fig:gradient} (color online) The curve surfaces
are the derivatives (in the unit of $\hbar\omega_\bot$) of the
ground-state energy along $(1, -1)$ direction versus $(\xi,
\Omega/\omega_\bot)$ for parameters $\sigma_1=0.2, \sigma_2=0.4,
\sigma_{12}=0.3$ (top panel) and
$\sigma_1=\sigma_2=\sigma_{12}=0.4$ (bottom panel) respectively.
The discontinuous steps precisely occur at boundaries separating
different phases illustrated in Fig.~\ref{fig:phim}. }
\end{figure}

Since we have the expressions of the energy levels near the ground
state, the phase diagram can be verified with the help of the
derivative of the ground-state energy. In Fig.~\ref{fig:gradient},
we plot the derivative of the ground-state energy along the $(1,
-1)$ direction versus $(\xi, \Omega/\omega_\bot)$,
\begin{equation}
F'(\xi,\Omega)=\frac{\partial\mathcal{E}_g}{\partial\xi}
  -\frac{\partial\mathcal{E}_g}{\partial(\Omega/\omega_\bot)},
\end{equation}
where $\mathcal{E}_g$ is the ground-state energy of the system.
One can see that discontinuous steps emerge in the $F'(\xi,
\Omega)$ surface (\ie, singularities for the derivative of
$F'(\xi, \Omega)$). This exhibits the boundaries separating
different phases as it is also an effective quantity to
characterize quantum phase transitions~\cite{Wu}. These boundaries
indeed concise with those given in Fig.~\ref{fig:phim}. As well
known, there are various quantities for characterizing quantum
phase transitions, such as the quantum
entanglement~\cite{Osterloh-etc} and the fidelity~\cite{Zanardi}.
The former is convenient if the reduced density matrix for paring
correlation is computable while the latter is convenient if the
ground state is expressible. In present case, the derivative of
the ground-state energy is the most convenient quantity for that
purpose.

The above discussion is based on the assumption of weak
interactions, which requires that
the interaction energy must be relatively much smaller
than kinetic and trapping  potential energies
for each component.
The former is of order
 $\sim\rho_ig_i$ (i=1,2), and the latter is of order $\sim\hbar\omega_i $,
 where $\rho_i$ is the particle density of the $i$th component in the $x$-$y$
 plane. The cross section of the $i$th component  is
$\pi a_i^2$ in the weak interaction limit. Thus, the density
$\rho_i$ is $~1/\pi a_i^2$.
So we can obtain  the effective coupling
constant $\sigma_i\ll1$ from the condition
$\rho_ig_i\ll\hbar\omega_i$.

If  the value of  $\sigma_i$  violates the condition of weak
interactions, the energy of the system will be minimized when many
components in the expansion of Eq.~(\ref{eq:wave}) contribute to
the wave function $\Psi_{i}$. In this case, the equations
(\ref{eq:gh}) is still valid. To calculate the expectation value
of $r_i^2$ and interaction Hamiltonian $H_{int}$ given in
Eq.~(\ref{eq:gh}), the numerical simulation and the averaged
vortex approximation~\cite{Ho} are efficient methods. Note that
two sets of vortex lattices may be formed in this case. The
structure of vortex lattices depends on the strength of
interactions as well as that of the gauge potentials. The presence
of gauge potentials can increase the diversity of vortex-lattice
structures, so it is worthwhile to further study these systems.

\section{Conclusion}
In summary, we have given a general description of the dark states
and found that $N-1$ degenerate dark states can be generated with
the help of $N$ laser fields coupling $N$-fold degenerate ground
states  with a common exited state. Interferences between two
waves with different frequencies can produce such radiation fields
that their  amplitudes are not only spatially dependent but also
time-dependent. They  can induce  vector and scalar gauge
potentials. As an example, we have considered a tripod system for
which one can obtain  a time-periodic gauge potential using a
specific laser field. We have given a configuration of laser field
that leads to a uniform $U(2)$ magnetic field, in which we have
discussed the features of Landau levels and studied the quantum
phase transitions of a rotating BEC using the derivative of the
ground-state energy.  We have shown that the presence of gauge
potentials will make the picture of rotating BECs to be much
richer.

This work is supported by NSFC No. 10225419 and No. 10674117.

\end{document}